\documentclass[a4paper, aps, prd,amsfonts, amssymb, amsmath, reprint, showkeys, nofootinbib, twoside,notitlepage,onecolumn]{revtex4-2}

\usepackage{amsmath,amstext}
\usepackage[T1]{fontenc}
\usepackage[colorlinks=true,linktocpage=true,linkcolor=blue,citecolor=blue,citecolor=red,urlcolor=blue]{hyperref}
\usepackage{amssymb}
\usepackage{graphicx}
\usepackage{ae,aecompl}
\usepackage{orcidlink}
\DeclareFontFamily{OT1}{pzc}{}
\DeclareFontShape{OT1}{pzc}{m}{it}{<-> s * [1.10] pzcmi7t}{}
\DeclareMathAlphabet{\mathpzc}{OT1}{pzc}{m}{it}

\usepackage{hyperref}
\usepackage{amsmath}
\usepackage{amssymb}
\usepackage{mathtools}
\usepackage{bm}
\usepackage{cleveref}
\usepackage{tensor}
\usepackage{braket}
\usepackage{enumitem}
\usepackage{mhchem}
\usepackage{amsthm}
\usepackage{nccmath}
\usepackage{mathrsfs}
\usepackage{color}

\newcommand{\taue}{\tau_\varepsilon}

\def\be{\begin{equation}}
\def\ee{\end{equation}}
\def\beq{\begin{eqnarray}}
\def\eeq{\end{eqnarray}}

\usepackage{xcolor}
\usepackage{ulem}
\definecolor{purple}{rgb}{0.8,0,0.6}

\theoremstyle{definition}

\theoremstyle{theorem}

\theoremstyle{corollary}

\begin{document}
\title{Hydrodynamics without a relaxation gap: memory effects, nonlocality, and superdiffusion}
\author{Lorenzo Gavassino$^1\orcidlink{0000-0002-6603-9253}$, Sukanya Mitra$^2\orcidlink{0000-0002-2401-957X}$, Rajeev Singh$^3$\orcidlink{0000-0001-5855-4039}}
\affiliation{$^1$Department of Applied Mathematics and Theoretical Physics, University of Cambridge, Wilberforce Road, Cambridge CB3 0WA, United Kingdom\\
$^2$School of Physical Sciences, National Institute of Science Education and Research,  An OCC of Homi Bhabha National Institute, Jatni 752050, India\\
$^3$Department of Physics, West University of Timisoara, Bulevardul Vasile Parvan 4, Timisoara 300223, Romania}

\begin{abstract}
By studying a simple model of relativistic particles propagating through a background medium with an energy-dependent relaxation time that is unbounded from above, we investigate how long-term memory obstructs the emergence of local hydrodynamics in systems with a gapless non-hydrodynamic sector. In the RTA matching frame, we show that the full gradient expansion is generically divergent in most flows, even for Fourier modes, and that any resummation necessarily retains an infinite set of slow non-hydrodynamic degrees of freedom. The divergence of the gradient expansion therefore reflects a more fundamental breakdown of hydrodynamic locality caused by persistent non-hydrodynamic memory. We finally show that sufficiently singular relaxation spectra can invalidate ordinary diffusion itself. In these regimes, the diffusivity diverges, and the late-time dynamics become superdiffusive, governed by intrinsically nonlocal constitutive relations.
\end{abstract} 
\maketitle

\section{Introduction}
\vspace{-0.4cm}

Hydrodynamics is commonly understood as a long-wavelength effective theory of many-body systems~\cite{landau6}. Its equations of motion are the local conservation laws, and its effective fields are the thermodynamic variables conjugate to the conserved charges. For example, consider an ensemble of particles propagating in an external medium. The relevant equation of motion is the particle-number conservation, $\partial_\mu J^\mu = 0$, where $J^\mu$ is the particle current, and the effective field is the fugacity $\varphi$ ($=\exp(\mu/T)$, with the exponential introduced for convenience). All information about the microscopic dynamics is encoded in a constitutive relation $J^\mu = J^\mu[\varphi]$, which is often organized as a derivative expansion as follows
\begin{equation}\label{gradientexpansion}
J^\mu = \mathcal{O}(1) + \mathcal{O}(\partial) + \mathcal{O}(\partial^2) + \cdots ,
\end{equation}
where successive terms are suppressed by increasing powers of the Knudsen number $\mathrm{Kn}=\lambda/L$ (with $\lambda$ the mean free path and $L$ the macroscopic gradient scale). The existence of such an expansion implicitly assumes a separation of length scales between the slow hydrodynamic evolution and the faster non-hydrodynamic relaxation processes.

This raises the question of what happens in systems whose non-hydrodynamic relaxation spectrum is gapless~\cite{GavassinoGapless:2024rck}, i.e.\ when the relaxation times of non-conserved degrees of freedom are unbounded from above. In such systems, there exist excitations that decay on arbitrarily long timescales, suggesting that the system may retain memory of the initial microscopic state indefinitely. In this scenario, the separation of scales are not unambiguously manifested. One therefore expects the emergence of local hydrodynamics to become obstructed, with the breakdown appearing through non-analytic constitutive behavior, divergent gradient expansions, or anomalous transport.

In this paper, we investigate this mechanism in a simple analytically tractable kinetic model of relativistic particles propagating through a background medium with an energy-dependent relaxation time that is unbounded above\footnote{An unbounded relaxation time typically emerges when the scattering cross section becomes asymptotically small in a certain energy sector, rendering the medium effectively transparent to particles in that regime. The resulting nearly free-streaming population gives rise to an infinite tower of non-hydrodynamic excitations whose lifetimes parametrically exceed those of the hydrodynamic modes \cite{GavassinoGapless:2024rck}. Such a tower of excitations is what we refer to as a gapless non-hydrodynamic spectrum.}. Our goal is to understand how the absence of a relaxation gap obstructs hydrodynamic locality. Working in a particular hydrodynamic frame (the RTA matching frame), we will show that the constitutive gradient expansion is generically divergent on all flows (not just in Bjorken-type flows \cite{HellerResurgence:2015dha,HellerBjorekn:2021oxl}, but also along the Fourier modes, where it is usually well-behaved \cite{Grozdanov:2019kge,HellerBounds2022ejw,GavassinoConvergence:2024xwf}) whenever the relaxation spectrum is gapless, and that any resummation necessarily retains infinitely many long-lived non-hydrodynamic degrees of freedom. We will also observe a similar mechanism at the spectral level: while a hydrodynamic dispersion relation $\omega(k)$ can formally be written down, its associated eigenfunction does not belong to the Hilbert space of finite-entropy excitations, and is therefore unphysical. Any physical excitation must instead involve a continuum of non-hydrodynamic modes, thereby reintroducing long-term memory into the dynamics. Thus, the divergence of the gradient expansion is a \textit{direct manifestation} of persistent non-hydrodynamic memory and the resulting breakdown of hydrodynamic locality. We will finally show that sufficiently singular relaxation spectra can invalidate ordinary diffusion itself. In this regime, the diffusivity diverges, and the late-time dynamics becomes superdiffusive, governed by intrinsically nonlocal constitutive behavior.

Recent studies of kinetic theories with energy-dependent relaxation times have emphasized the non-analytic structure of retarded correlation functions~\cite{Brants:2024wrx}. Our analysis is complementary, as it reformulates the same obstruction as the impossibility of reducing the dynamics to local constitutive relations for conserved variables alone.

In this paper, we work in the Minkowski space with a mostly positive metric signature \((-,+,+,+)\), and adopt the natural units \(c=\hbar=k_B=1\). The Greek indices run over spacetime components (\(0\) to \(3\)), while the Latin indices run over spatial components (\(1\) to \(3\)).

\vspace{-0.3cm}
\section{Our kinetic toy-model}
\vspace{-0.3cm}

Throughout the paper, we work with a linearized kinetic model of massless particles propagating in an external medium, with a single conserved charge and no particle-particle collisions. Below is a brief overview of this model.

\vspace{-0.3cm}
\subsection{The Boltzmann equation in the relaxation-time approximation}
\vspace{-0.3cm}

We consider a dilute gas of massless particles, each with energy $\varepsilon = \sqrt{p^j p_j}$ and velocity $v^j = \partial \varepsilon / \partial p_j = p^j / \varepsilon$, where $p_j$ is the particle momentum. The particles propagate through a background medium at rest, and scatter off impurities, but do not interact with each other. Assuming that the medium acts as a Markovian bath at inverse temperature $\beta$, and that the scattering rate depends on the particle energy, the evolution of the one-particle distribution function $f(x^\mu, p^j)$ can be modeled by the Boltzmann transport equation in the Relaxation-Time Approximation (RTA\footnote{In a dilute gas whose collisions are dominated by binary particle--particle scattering, RTA is usually only a rough approximation to the full Boltzmann collision integral. In the present setting, however, where scattering is mediated by an external medium, it is much better justified. The loss term $-f/\tau_\varepsilon$ describes the usual attenuation of a beam across randomly distributed scattering centers. If collisions with the medium fully randomize the outgoing state, then the gain term can depend on $f$ only through an overall normalization fixed by particle-number conservation, and must therefore take the form $\varphi[f]G(\varepsilon)$, with $G$ independent of $f$. Requiring the collision term to vanish in equilibrium forces $G(\varepsilon)\propto e^{-\beta\varepsilon}/\taue$, yielding precisely the RTA collision operator adopted here.}) with an energy-dependent relaxation time $\tau_\varepsilon$ in the following manner:
\begin{equation}\label{boltzmann}
(\partial_t + v^j \partial_j) f = \frac{\varphi\, e^{ - \beta \varepsilon} - f}{\tau_\varepsilon} \, .
\end{equation}
Here, the field $\varphi(x^\mu)$ is a functional of $f$, determined by the requirement that the particle-number current, given by the following moment integral over the single-particle distribution function $f$ as
\begin{equation}\label{current}
J^\mu = \int \frac{d^3 p}{(2\pi)^3}
\begin{bmatrix}
1 \\
v^j
\end{bmatrix}
f
\end{equation}
must be conserved, i.e.\ $\partial_\mu J^\mu = 0$. This leads to the matching condition
\begin{equation}\label{matching}
\int \frac{d^3 p}{(2\pi)^3} \frac{\varphi\, e^{ - \beta \varepsilon} - f}{\tau_\varepsilon}=0~,\qquad \qquad \Longleftrightarrow  \qquad \qquad
\varphi = \frac{\displaystyle \int \frac{d^3 p}{(2\pi)^3} \frac{f}{\tau_\varepsilon}}{\displaystyle \int \frac{d^3 p}{(2\pi)^3} \frac{e^{-\beta \varepsilon}}{\tau_\varepsilon}} \, .
\end{equation}
With this choice of matching, Eq.~\eqref{boltzmann} is linear in $f$, consistent with the absence of particle-particle interactions.

\vspace{-0.3cm}
\subsection{Thermodynamic consistency}
\vspace{-0.3cm}

Since the medium acts as a heat bath, the energy and momentum of the gas are not conserved, and its entropy is not necessarily non-decreasing. However, the Helmholtz free energy is expected to be non-increasing. In kinetic theory, the free-energy current of a dilute gas takes the form,
\begin{equation}\label{Helmholtz}
\mathcal{F}^\mu = \frac{1}{\beta} \int \frac{d^3 p}{(2\pi)^3}
\begin{bmatrix}
1 \\
v^j
\end{bmatrix}
f \, (\beta \varepsilon + \ln f - 1) \, .
\end{equation}
Using \eqref{boltzmann}, together with the matching condition \eqref{matching}, one finds that the divergence of the free energy current reads
\begin{equation}
\partial_\mu \mathcal{F}^\mu = -\frac{1}{\beta} \int \frac{d^3 p}{(2\pi)^3 \tau_\varepsilon} (\varphi\, e^{ - \beta \varepsilon} - f)
\ln \left( \frac{\varphi\, e^{ - \beta \varepsilon}}{f} \right) \, ,
\end{equation}
which is non-positive provided that $\tau_\varepsilon > 0$. Hence, the model satisfies a thermodynamic second-law-type inequality, and the total free energy $F = \int d^3 x \, \mathcal{F}^0$ acts as a Lyapunov functional.

The minimum of $F$ (at fixed particle number) is attained when $f =\varphi\, e^{ - \beta \varepsilon}$, corresponding to the Maxwell-J\"uttner distribution of relativistic particles. This establishes the thermodynamic consistency of the model. It also automatically implies that the causality and covariant stability of the kinetic evolution is preserved \cite{GavassinoCausality2021,GavassinoDistrubing:2026klp}.

\section{Divergence of the gradient expansion in the RTA matching frame}
\vspace{-0.3cm}

Having constructed a self-consistent kinetic model for particle diffusion in a medium, we now turn to the behavior of the gradient expansion \eqref{gradientexpansion} in a generic state. To this end, one must specify a field $\varphi(x^\mu)$ that plays the role of the fugacity out of equilibrium, i.e.\ one must fix a hydrodynamic frame \cite{Bemfica:2017wps,Kovtun:2019hdm}. 
A natural choice is provided by the parameter $\varphi$ entering the Boltzmann equation \eqref{boltzmann}, which is determined by the matching condition \eqref{matching}. In equilibrium, this quantity reduces to the standard fugacity. For an energy-dependent relaxation time, however, the resulting hydrodynamic frame does not coincide with the usual frames (e.g., Eckart \cite{Eckart40}, or Landau \cite{landau6}), but instead defines a distinct ``RTA matching frame''. In this frame, the gradient expansion can be computed straightforwardly to arbitrary order, allowing us to study its convergence rigorously.

\vspace{-0.3cm}
\subsection{Formal derivation}
\vspace{-0.3cm}

Fix a phase-space point $(x^\mu,p^j)$. We define the characteristic curve
$\{x^\mu(\xi),p^j(\xi)\}=\{x^\mu-\tau_\varepsilon \xi\, p^\mu/\varepsilon,\, p^j\}$,
where $p^\mu=(\varepsilon,p^j)$ is the particle four-momentum, and $\xi$ denotes the optical depth. Along this trajectory, the Boltzmann equation \eqref{boltzmann} reduces to the ordinary differential equation 
\(
df/d\xi - f = - \varphi\, e^{-\beta \varepsilon}\).
Integrating from $\xi=0$ to $\xi=+\infty$, we obtain the following solution:
\begin{equation}\label{integraloff}
f(x^\mu,p^j)=e^{-\beta\varepsilon}\int_0^{+\infty} d\xi \, e^{-\xi} \, \varphi(x^\mu - \xi \tau_\varepsilon p^\mu/\varepsilon) \, .
\end{equation}
Now, treating $\varphi$ as a function of $\xi$, we expand it in a Taylor series around $\xi=0$ in the following manner,
\begin{equation}\label{Taylor}
\varphi(\xi)=\sum_{n=0}^{\infty} \frac{\xi^n}{n!}\left(\frac{d}{d\xi}\right)^n\varphi(0)~,
\end{equation}
where $\frac{d}{d\xi}=-\tau_{\varepsilon}\frac{p^{\mu}}{\varepsilon}\partial_{\mu}$ defines the derivative along the characteristic line. Plugging \eqref{Taylor} into \eqref{integraloff} and
assuming that the series can be interchanged with the integral, we obtain
\begin{equation}\label{seriesf}
f=e^{-\beta\varepsilon} \sum_{n=0}^{+\infty} \left(-\tau_\varepsilon \frac{p^\nu}{\varepsilon}\partial_\nu \right)^n \varphi \, .
\end{equation}
Substituting this expression into \eqref{current}, and again assuming that the series can be interchanged with the momentum integral, we obtain an infinite-order constitutive relation for $J^\mu$ in terms of $\varphi$:
\begin{equation}\label{expansionna}
J^\mu[\varphi]= \sum_{n=0}^{+\infty} (-1)^n \partial_{\nu_1}\cdots\partial_{\nu_n} \varphi\int \frac{d^3 p}{(2\pi)^3}
\frac{p^\mu}{\varepsilon}
\frac{\tau_\varepsilon p^{\nu_1}}{\varepsilon} \cdots \frac{\tau_\varepsilon p^{\nu_n}}{\varepsilon}\, e^{-\beta\varepsilon} \, ,
\end{equation}
which may be understood as a formal gradient expansion, and will be the main object of study in the following.

Let us note that, at a given order, the expansion \eqref{expansionna} contains the same number of time and space derivatives. In general, time derivatives may be eliminated order by order using the conservation law $\partial_\mu J^\mu=0$ \cite{GavassinoInitialData:2026xjw}. The choice of how many time derivatives to retain is part of the hydrodynamic frame choice: at one extreme, the BDNK (Bemfica-Disconzi-Noronha-Kovtun) formulation retains time and space derivatives on equal footing~\cite{Bemfica:2017wps,Kovtun:2019hdm}, while at the other extreme, the density frame eliminates all time derivatives from the constitutive relations~\cite{Armas:2020mpr,Basar:2024qxd,Bhambure:2024axa,Bhambure:2024gnf}. Here, we are following the spirit of the BDNK formulation and keeping both the temporal and spatial derivatives.

\vspace{-0.3cm}
\subsection{Quick estimates}
\vspace{-0.3cm}

To determine under which conditions the series \eqref{expansionna} converges, let us perform a rough order-of-magnitude estimate. Suppose that $\partial^n \sim a_n / L^n$, where $L$ is the characteristic hydrodynamic gradient length scale, and $a_n$ is the dimensionless coefficients encoding the growth of $n^{th}$ derivative for a given density profile. For massless particles, we have $p^\nu / \varepsilon \sim 1$, so that \eqref{expansionna} becomes
\vspace{-0.3cm}
\begin{equation}
J^\mu[\varphi] \sim \varphi \sum_{n=0}^{+\infty} \int \frac{d^3 p}{(2\pi)^3}
\left( \frac{\tau_\varepsilon}{L} \right)^n a_n  \, e^{ - \beta \varepsilon} \, .
\end{equation}
For a sinusoidal plane wave, one has $a_n \sim 1$, and the series reduces to a geometric series with ratio $\tau_\varepsilon / L$. It therefore converges only if $\tau_\varepsilon < L$ for all $\varepsilon$ (up to a set of measure zero). In this case, the gradient expansion converges provided that the Knudsen number $\mathrm{Kn} = \sup(\tau_\varepsilon)/L$ is smaller than unity. Conversely, if $\tau_\varepsilon$ is unbounded from above, i.e.\ if the system is gapless, the series necessarily diverges, even for plane-wave configurations.

For more general density profiles, which can be decomposed into superpositions of all Fourier modes, convergence is even more restrictive. For instance, for a Lorentzian profile one finds $a_n \sim n!$, and the series diverges for any value of $\tau_\varepsilon / L$. One might attempt to avoid this divergence by considering profiles that are polynomial in space, so that $a_n = 0$ beyond some finite order. However, since the expansion \eqref{expansionna} also involves time derivatives, this would require a similar truncation in time as well, which is not satisfied by the dynamics in general.

\vspace{-0.3cm}
\subsection{A more precise example}\label{amoreprecise}
\vspace{-0.3cm}

The estimates of the previous subsection were heuristic. Let us now consider an explicit example in which the analysis can be carried out exactly. We consider a plane wave $\propto e^{ikx - i\omega t}$, and assume that the relaxation time follows a power law in the energy, $\tau_\varepsilon = g\, \varepsilon^\gamma$, with $g$ and $\gamma$ constant. Then, the zeroth component of \eqref{expansionna} becomes
\begin{equation}\label{expansionna2}
J^0[\varphi]= \varphi \sum_{n=0}^{+\infty} g^n \int \frac{d^3 p}{(2\pi)^3} \, \left(i\omega - i k v^1\right)^n
\varepsilon^{n\gamma} \, e^{-\beta\varepsilon} \, .
\end{equation}
We immediately see that, if $\gamma<0$, all momentum integrals with $n\gamma \le -3$ are infrared-divergent due to a low-energy singularity. The series therefore is ill-defined in this case, and we may thus restrict attention to $\gamma \ge 0$. Then, expressing the integrals in spherical coordinates, and performing the angular and energy integrals separately, we obtain
\begin{equation}
\begin{split}
J^0[\varphi] &= \varphi \sum_{n=0}^{+\infty} (gik)^n \int_{-1}^1 \frac{dv^1}{2} \left(\frac{\omega}{k}-v^1\right)^n \int_0^{+\infty} \frac{d\varepsilon }{2\pi^2} 
\varepsilon^{n\gamma+2} \, e^{-\beta\varepsilon} \\
&= \varphi \sum_{n=0}^{+\infty} \Gamma(3+n\gamma) (gik)^n \, \frac{\left(\frac{\omega}{k}+1\right)^{n+1}-\left(\frac{\omega}{k}-1\right)^{n+1}}{(2\pi)^2 (n+1)\, \beta^{3+n\gamma}} \, .
\end{split}
\end{equation}
For $\gamma>0$, the factor $\Gamma(3+n\gamma)$ grows super-exponentially with $n$, so the terms of the series do not tend to zero and the sum diverges, irrespective of any oscillatory factors.

The only remaining case is $\gamma=0$. Then, up to subleading prefactors such as $n+1$, the series reduces to the difference of two geometric series with common ratios $gik(\frac{\omega}{k}+1)$ and $gik(\frac{\omega}{k}-1)$. In the long-wavelength regime, where $\omega \sim -i k^2$ and thus $\omega/k \to 0$, convergence requires $|gk|<1$, which is equivalent to the Knudsen number condition $\tau/L < 1$, since $k$ denotes the spatial gradients indicating the inverse of macroscopic length scale.

In summary, this example confirms the heuristic argument of the previous subsection: whenever the relaxation time is unbounded from above (here, for $\gamma\neq 0$), the gradient expansion diverges or is ill-defined. For $\gamma<0$ (i.e. large relaxation times at low energies), the series coefficients themselves diverge due to the infrared blow-up of the momentum integrals, while for $\gamma>0$ (i.e. large relaxation times at high energies) the series diverges due to the large-order growth of the expansion coefficients. By contrast, in the gapped case $\gamma=0$, the gradient expansion may converge, provided that the gradients are sufficiently small (similar results hold for the components $J^k$).

\vspace{-0.3cm}
\subsection{Regularizing the constitutive relation requires reintroducing non-hydrodynamic modes}
\vspace{-0.3cm}

To elucidate the origin of the divergence, let us reconsider the integral in \eqref{integraloff}, which, as discussed, retraces the past history of the beam of particles with momentum $p^\mu$ crossing the event $x^\mu$ (with integration variable $\xi = (t - t')/\tau_\varepsilon$). For concreteness, assume that $\varphi$ exhibits a plane-wave dependence of the form $e^{ik_\mu x^\mu}$. In this case, the integrand behaves as $e^{-\xi\left(1 + \tau_\varepsilon \, i k_\mu p^\mu / \varepsilon \right)}$, meaning that the integral converges if and only if
\begin{equation}\label{convergencecondit}
\mathfrak{Re}\!\left(1 + \tau_\varepsilon \, i k_\mu p^\mu / \varepsilon \right) > 0.
\end{equation}
For a Fourier mode ($k_j \,{\in}\, \mathbb{R}$), the above condition reduces to $\mathfrak{Im}\,\omega\, {>}\, {-}1/\tau_\varepsilon$, which implies that the construction \eqref{integraloff} yields a finite result only if the mode decays on timescales longer than the relaxation time $\tau_\varepsilon$. Hence, if $\tau_\varepsilon$ is unbounded from above, there is always an energy sector where the distribution function \eqref{integraloff} diverges, implying that all subsequent manipulations (expanding $f$ in series and exchanging the sum with the momentum integral) are no longer allowed.

How, then, should one regularize the constitutive relation? For those four-momenta where \eqref{convergencecondit} is violated, the integral representation \eqref{integraloff} can no longer be extended to $\xi \to +\infty$, i.e.\ to the infinite past of the beam's worldline. Instead, the integration must be truncated at a finite value $\xi_0$, corresponding to an initial time at which $f$ is specified. One then obtains,
\begin{equation}\label{integraloff2}
f(x^\mu,p^j)=e^{-\xi_0}f\!\left(x^\mu - \xi_0 \tau_\varepsilon p^\mu/\varepsilon,p^j\right)
+e^{-\beta\varepsilon}\int_0^{\xi_0} d\xi \, e^{-\xi} \,
\varphi\!\left(x^\mu - \xi \tau_\varepsilon p^\mu/\varepsilon\right) \, .
\end{equation}
This expression shows that $J^\mu$ can no longer be written solely as a functional of $\varphi$. Instead, the solution acquires an explicit dependence on the initial data through the cutoff $\xi_0$, reflecting the persistence of long-term memory. Thus, constructing a finite constitutive relation requires reintroducing the non-hydrodynamic modes whose relaxation times exceed the hydrodynamic timescale, and therefore cannot be integrated out~\cite{Bhattacharyya:2024tfj,Bhattacharyya:2024ohn}. In a gapless system, this inevitably restores an infinite set of slow degrees of freedom to the description.

\vspace{-0.3cm}
\section{Memory-related pathologies of the hydrodynamic mode}
\vspace{-0.3cm}

We have shown that, in a particular hydrodynamic frame, the constitutive gradient expansion becomes divergent when the relaxation spectrum is gapless, and that restoring a finite constitutive relation requires reintroducing infinitely many long-lived non-hydrodynamic degrees of freedom. We now reformulate the same obstruction directly at the level of the excitation spectrum, where it can be understood independently of any hydrodynamic frame choice.

\vspace{-0.3cm}
\subsection{Formulation of the problem}
\vspace{-0.3cm}

Fix the global equilibrium state $e^{-\beta\varepsilon}$, and decompose the out-of-equilibrium distribution as $f = e^{-\beta\varepsilon} + \sqrt{e^{-\beta\varepsilon}}\,\psi$, where $\psi$ is some perturbative correction. Then, to quadratic order in $\psi$, the Helmholtz free energy \eqref{Helmholtz} reads
\begin{equation}
\mathcal{F}^0[f]
=
\mathcal{F}^0[e^{-\beta\varepsilon}]
+
\dfrac{1}{2\beta}
\int \frac{d^3 p}{(2\pi)^3} |\psi|^2
+
\mathcal{O}(\psi^3) \, .
\end{equation}
Thus, fluctuations with finite free energy correspond to $\psi \in L^2(\mathbb{R}^3)$, which defines the natural Hilbert space for linearized perturbations \cite{GavassinoGapless:2024rck,GavassinoConvergence:2024xwf,GavassinoDistrubing:2026klp}, equipped with the standard (Onsager-type) inner product $(\psi,\tilde{\psi}) = \int \frac{d^3 p}{(2\pi)^3} \psi^* \tilde{\psi}$. 
Within this space, the Boltzmann equation \eqref{boltzmann} for $\psi \propto e^{ikx-i\omega t}$ reduces to
\begin{equation}\label{eigenvalueproblem}
\left(\dfrac{1}{\tau_\varepsilon} + ik v^1 \right)\psi
-
\dfrac{e^{-\beta\varepsilon/2}}{\tau_\varepsilon}
\dfrac{(e^{-\beta\varepsilon/2}/\tau_\varepsilon,\psi)}
{(e^{-\beta\varepsilon/2}/\tau_\varepsilon,e^{-\beta\varepsilon/2})}
=
i\omega \psi \, ,
\end{equation}
which may be viewed as an eigenvalue problem $L_{ik}\psi = i\omega \psi$, where $L_{ik}$ is a linear operator on $L^2(\mathbb{R}^3)$. This operator is the sum of the multiplication operator $\left(1/\tau_\varepsilon + ik v^1\right)$, whose spectrum coincides with its essential range and is given by the continuous band
\begin{equation}\label{essentialspetrum}
i\omega
\in
\bigcup_{p^j\in \mathbb{R}^3}
\left(
\dfrac{1}{\tau_\varepsilon} + ik v^1
\right) \, ,
\end{equation}
minus a rank-one operator $\zeta(\zeta,\cdot)$, with
\begin{equation}
\zeta
=
\dfrac{1}{\sqrt{(e^{-\beta\varepsilon/2}/\tau_\varepsilon,e^{-\beta\varepsilon/2})}}
\dfrac{e^{-\beta\varepsilon/2}}{\tau_\varepsilon} \, .
\end{equation}
Since the latter is compact, it does not modify the essential spectrum \eqref{essentialspetrum}.

Our task is therefore to identify the solutions of \eqref{eigenvalueproblem} that belong to the Hilbert space, i.e.\ are square-integrable. Such solutions are physically admissible, since they possess finite free energy and finite moments. Indeed, the $n$-th energy moment is $(\varepsilon^n e^{-\beta\varepsilon/2},\psi)
\leq
||\varepsilon^n e^{-\beta\varepsilon/2}||_{L^2}
\, ||\psi||_{L^2}
<
\infty$,
by the Cauchy-Schwarz inequality. Moreover, from the perspective of spectral theory, such solutions correspond to genuine eigenvalues of the evolution operator, and therefore appear as pole-like contributions in retarded correlators, unlike the essential spectrum, which manifests itself as a non-analytic region in the complex plane~\cite{Brants:2024wrx}.

If among these solutions there exists an $L^2$ eigenfunction whose eigenvalue satisfies $\omega(k)\to0$ as $k\to0$, and whose eigenfunction reduces to $e^{-\beta\varepsilon/2}$ at $k=0$ (corresponding to a neighboring equilibrium state generated by a global shift of the conserved charge), then such a branch may be identified as the hydrodynamic mode of the system.

\vspace{-0.3cm}
\subsection{General structure of the solution}
\vspace{-0.3cm}

Rearranging \eqref{eigenvalueproblem}, we obtain $(\frac{1}{\tau_\varepsilon} - i\omega + ik v^1)\psi = (\zeta,\psi)\,\zeta$.
The case $(\zeta,\psi)=0$ can be excluded, as it would require $\psi$ to be supported on the set of $p^j$ where $v^1=(i\omega\tau_\varepsilon-1)/(ik\tau_\varepsilon)$, which is generically a set of measure zero in $\mathbb{R}^3$, implying that $\psi\,{=}\,0$ in $L^2(\mathbb{R}^3)$. By linearity, we may thus fix the normalization $(\zeta,\psi)=\sqrt{(e^{-\beta\varepsilon/2}/\tau_\varepsilon,e^{-\beta\varepsilon/2})}$, 
yielding
\begin{equation}\label{psipsipsi}
\psi = \dfrac{e^{-\beta\varepsilon/2}}{1 - (i\omega - ik v^1)\tau_\varepsilon} \, .
\end{equation}
Plugging this expression for $\psi$ into the normalization equation mentioned above then gives an equation for $\omega$:
\begin{equation}\label{determinantequation}
\int_0^{\infty} d\varepsilon\left[ \dfrac{\varepsilon^2  e^{-\beta\varepsilon} }{\tau_\varepsilon} \right]~\int_{-1}^{1}\dfrac{dv^1}{2} \, \left[ 1-\dfrac{1}{1-\left\{\tau_\varepsilon(i\omega-ikv^1)\right\}} \right]=0 \, .
\end{equation}
The task is now to study the analyticity of the dispersion relation \eqref{determinantequation} in the complex $(\omega,k)$ plane.

\subsection{Non-analyticity of the dispersion relation}\label{singularities}

\begin{figure}[b!]
    \centering
\includegraphics[width=0.4\linewidth]{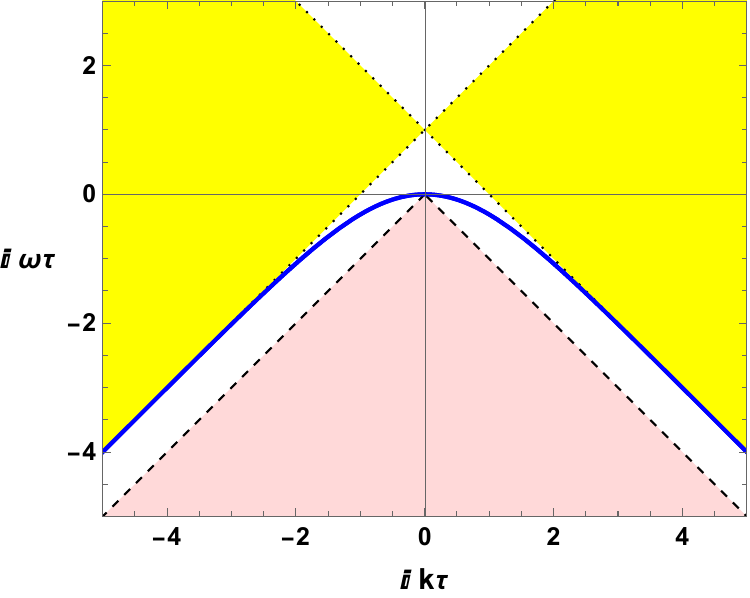}\hspace{0.08\linewidth}
\includegraphics[width=0.39\linewidth]{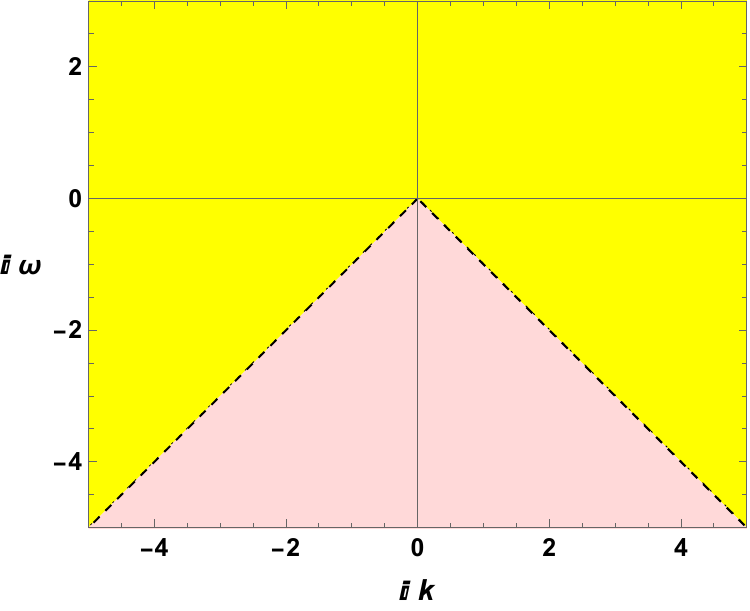}
\caption{Excitation spectrum of RTA kinetic theory for imaginary wavenumbers. For the distribution function to be non-singular, the dispersion relation of the hydrodynamic mode (blue line) must lie outside the essential spectrum \eqref{essentialspetrum} (yellow region). In addition, covariant stability requires $\mathfrak{Im}\,\omega \leq |\mathfrak{Im}\,k|$, excluding the red region. 
Left panel: for a constant relaxation time $\tau$, a finite gap separates the essential spectrum from the unstable region, allowing for a smooth diffusive dispersion relation to exist in between. In the RTA case, such dispersion relation is known analytically \cite{RomatschkeCutsandPoles:2015gic,Bajec:2025dqm}: $i\omega\tau = 1 - ik\tau/\tanh(ik\tau)=1-k\tau~ \text{cot}(k\tau)$. 
Right panel: when $\tau_\varepsilon$ grows as a power law in the energy, the essential spectrum becomes gapless and touches the unstable region along the line $i\omega = -|ik|$. Any stable eigenvalue with a regular eigenfunction $\psi$ must then lie on this line, which is non-smooth at the origin. Consequently, the hydrodynamic dispersion relation $\omega(k)$ (if it exists) cannot be analytic.}
    \label{fig:GappedVsGapless}
\end{figure}

Let us fix $ik \in \mathbb{R}$, so that the operator $L_{ik}$ in \eqref{eigenvalueproblem} is self-adjoint, and hence $i\omega$, being an eigenvalue, must also be real. 
We now ask for which values of $i\omega \in \mathbb{R}$ the candidate eigenstate \eqref{psipsipsi} belongs to $L^2(\mathbb{R}^3)$. The denominator vanishes for some $p^j$ whenever $i\omega$ lies in the interior of the essential spectrum \eqref{essentialspetrum}. In this case, $\psi$ fails to be square-integrable due to a divergence in a neighborhood of the critical surface $v^1=(i\omega\tau_\varepsilon-1)/(ik\tau_\varepsilon)\equiv v^1_c(\varepsilon)$ (assuming $k\taue \neq 0$) since
\begin{equation}
||\psi||_{L^2}^2 = \int \dfrac{d^3p}{(2\pi)^3} \dfrac{e^{-\beta\varepsilon}}{|1 - (i\omega {-} ik v^1)\tau_\varepsilon|^2}
= \int_0^\infty \dfrac{d\varepsilon}{(2\pi)^2} \dfrac{\varepsilon^2 e^{-\beta\varepsilon}}{|k\tau_\varepsilon|^2} \int_{-1}^1 \dfrac{dv^1}{|v^1{-}v^1_c(\varepsilon)|^2} = \infty  \qquad \binom{\text{ if }v^1_c(\varepsilon)\in [-1,1]\,}{\text{ on an interval of }\varepsilon} .
\end{equation}
One might attempt to relax the requirement that $\psi$ belong to $L^2$. However, for $f = e^{-\beta\varepsilon} + e^{-\beta\varepsilon/2}\,\psi$ to be a physically acceptable distribution function, it should be at least $L^1$. Unfortunately, if $i\omega$ lies in the interior of the essential spectrum, the singularity of $\psi$ is not integrable even in the $L^1$ sense: 
\begin{equation}
|| \sqrt{e^{-\beta\varepsilon}}\,\psi||_{L^1} = \int \dfrac{d^3p}{(2\pi)^3} \dfrac{e^{-\beta\varepsilon}}{|1 - (i\omega {-} ik v^1)\tau_\varepsilon|}
= \int_0^\infty \dfrac{d\varepsilon}{(2\pi)^2} \dfrac{\varepsilon^2 e^{-\beta\varepsilon}}{|k\tau_\varepsilon|} \int_{-1}^1 \dfrac{dv^1}{|v^1{-}v^1_c(\varepsilon)|} = \infty  \quad \binom{\text{ if }v^1_c(\varepsilon)\in [-1,1]\,}{\text{ on an interval of }\varepsilon} .
\end{equation}
Hence, $i\omega$ must lie outside (or on the boundary of) the essential spectrum. In a gapless system, however, there exists no \textit{smooth} function $\omega(k): i\mathbb{R}\to i\mathbb{R}$ that passes through the origin while remaining outside the essential spectrum. In fact, for imaginary $k$, the latter fills a continuous region with lower boundary $i\omega = -|ik|$ (see figure \ref{fig:GappedVsGapless}), and the dispersion relation therefore must have a discontinuous derivative at $k=0$.

The above argument shows that, in a gapless system, the hydrodynamic dispersion relation admits no smooth continuation to imaginary $k$, and therefore cannot be analytic at the origin. The same obstruction also appears directly in equation \eqref{determinantequation}. Upon complexifying $i\omega$ and $ik$, the angular integral over $v^1$ can be evaluated to give
\begin{equation}\label{determinantequationATANH}
F(\omega,k)\equiv \int_{0}^\infty d\varepsilon \,
\dfrac{\varepsilon^2 e^{-\beta\varepsilon}}{\tau_\varepsilon}
\left[
1-
\dfrac{1}{ik\tau_\varepsilon}
\text{arctanh}
\left(
\dfrac{ik\tau_\varepsilon}{1-i\omega\tau_\varepsilon}
\right)
\right]
=0 \, .
\end{equation}
In a gapped system, the argument of $\mathrm{arctanh}$ remains bounded away from the logarithmic branch points $\pm1$ for sufficiently small $\omega$ and $k$. Hence, $F(\omega,k)$ remains holomorphic in a neighborhood of the origin, and the implicit function theorem can be applied to obtain a regular dispersion relation $\omega(k)$.
In a gapless system, however, $\tau_\varepsilon$ becomes arbitrarily large, so that for any nonzero $\omega$ and $k$ the argument of $\mathrm{arctanh}$ reaches the singular values $\pm1$ at sufficiently large or small energies (depending on the behavior of $\tau_\varepsilon$). The function $F$ therefore generically ceases to be holomorphic near $(\omega,k)\,{=}\,(0,0)$, implying that the hydrodynamic dispersion relation $\omega(k)$ is non-analytic at $k\,{=}\,0$.

\subsection{Long-term memory}\label{longtermmemory}

\begin{figure}[b!]
    \centering
\includegraphics[width=0.44\linewidth]{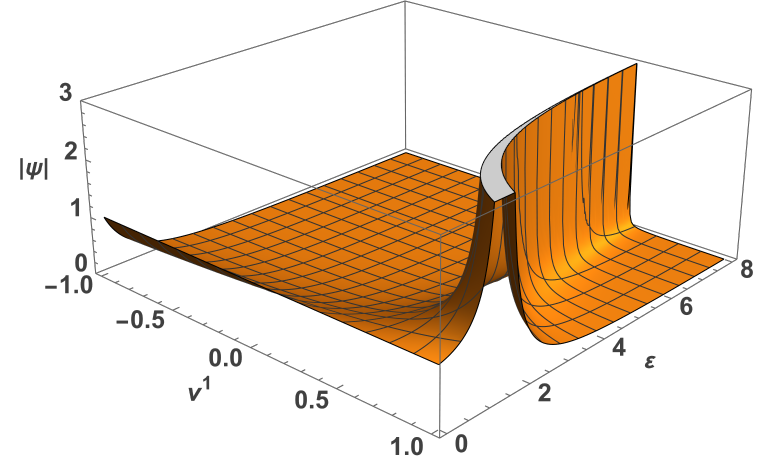}\hspace{0.08\linewidth}
\includegraphics[width=0.44\linewidth]{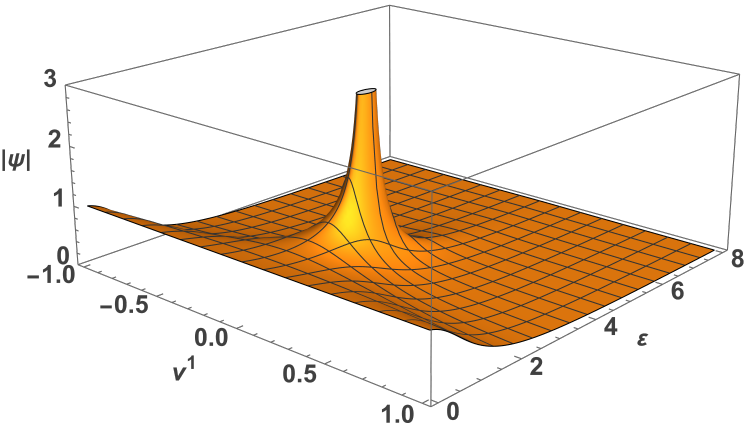}
\caption{
Singular behavior of the candidate hydrodynamic distribution function $|\psi|$ in a gapless system, shown as a function of $(\varepsilon,v^1)$ for the model $\beta=1$ and $\tau_\varepsilon=\varepsilon$. Left panel: for imaginary wavenumber ($\omega=i/3$, $k=i$), the singularity extends along a codimension-one curve, $v^1=1/3+1/\varepsilon$. This produces a non-integrable divergence in both $L^1$ and $L^2$. Right panel: for real wavenumber ($\omega=-i/3$, $k=1$), the singularity collapses to an isolated codimension-two point at $(\varepsilon,v^1)=(3,0)$. The distribution function remains non-square-integrable, but becomes locally integrable, allowing the dispersion relation to remain well defined despite the singularity.
Note that the values of $\omega$ and $k$ used in these plots are chosen only to illustrate the singularity structure, and do not necessarily solve \eqref{determinantequation}.}
    \label{fig:RealvsImagk}
\end{figure}

Let us now fix $k\in \mathbb{R}\setminus\{0\}$, corresponding to a non-equilibrium Fourier mode, which is the relevant building block for studying the dynamics of localized perturbations. From equation \eqref{determinantequation}, one sees that if $i\omega$ is a mode of the system, then $(i\omega)^*$ is also a mode for the same $k$. Assuming that the hydrodynamic mode is unique (since there is only one conserved charge), it follows that $i\omega\in\mathbb{R}$, consistent with purely diffusive behavior. Under these assumptions, the absolute value of \eqref{psipsipsi} reads
\begin{equation}\label{singularweaker}
|\psi|
=
\dfrac{e^{-\beta\varepsilon/2}}
{\sqrt{(1-i\omega\tau_\varepsilon)^2 +(k\tau_\varepsilon v^1)^2}} \, .
\end{equation}
This distribution function still develops a singularity in momentum space, located at $\tau_\varepsilon=(i\omega)^{-1}$ and $v^1=0$. However, unlike the imaginary-wavenumber case, the singular set now has codimension two rather than one. In the $(\varepsilon,v^1)$ plane, the divergence is therefore localized at an isolated point rather than along a curve (see figure \ref{fig:RealvsImagk}). Assuming that $v_c^1(\varepsilon)=(i\omega\tau_\varepsilon-1)/(k\tau_\varepsilon)$ has non-vanishing derivative at its zero, the relevant norms behave as
\vspace{-0.1cm}
\begin{equation}\label{mildersingularity}
\begin{split}
||\psi||_{L^2}^2
&=
\int \dfrac{d^3p}{(2\pi)^3}
\dfrac{e^{-\beta\varepsilon}}
{(1-i\omega\tau_\varepsilon)^2 +(k\tau_\varepsilon v^1)^2}
=\infty \, , \\
|| \sqrt{e^{-\beta\varepsilon}}\,\psi||_{L^1}
&=
\int \dfrac{d^3p}{(2\pi)^3}
\dfrac{e^{-\beta\varepsilon}}
{\sqrt{(1-i\omega\tau_\varepsilon)^2 +(k\tau_\varepsilon v^1)^2}}
<\infty \, .
\end{split}
\end{equation}
Hence, while the candidate hydrodynamic mode still has infinite free energy, it defines a generalized distribution with finite moments. In particular, equation \eqref{determinantequation} remains well defined, and therefore still determines a dispersion relation $\omega(k)$, even though this function is non-analytic at $k=0$.

Nevertheless, any physical realization of the mode must regularize the singularity. Instead of \eqref{singularweaker}, any realistic experimental setup would always produce a smooth approximation of the form
\vspace{-0.1cm}
\begin{equation}\label{experiment}
|\psi_{\mathrm{experiment}}|
\sim
\dfrac{e^{-\beta\varepsilon/2}}
{\sqrt{(1-i\omega\tau_\varepsilon)^2 +(k\tau_\varepsilon v^1)^2+\delta^2}} \, ,
\end{equation}
with $\delta>0$ determined by the preparation of the initial state. Such a regularization introduces non-hydrodynamic excitations concentrated near the resonant energy band $\tau_\varepsilon=(i\omega)^{-1}$. These modes relax on the same timescale as the hydrodynamic mode itself, so their relative contribution never disappears dynamically. The system therefore retains a permanent memory of the non-hydrodynamic sector, and the late-time evolution cannot be reduced to a purely hydrodynamic description.

This is the same mechanism encountered in the previous section. The formal hydrodynamic solution is singular, and any physical regularization necessarily restores long-lived non-hydrodynamic degrees of freedom whose dynamics cannot be integrated out.

\section{Possibility of superdiffusion}
\label{sec:superdiffusion}

As discussed in subsection \ref{amoreprecise}, not only the full gradient expansion, but also individual transport coefficients may diverge. This possibility is excluded in gapped kinetic theories, where the dispersion relation $\omega(k)$ is analytic in a neighbourhood of $k=0$ \cite{GavassinoConvergence:2024xwf}, but can arise in gapless systems. In particularly extreme situations, even the diffusivity coefficient $\mathfrak{D}$ (defined through the formal expansion $\omega = -i \mathfrak{D} k^2 + \dots$) may diverge; in such cases, one speaks of \textit{superdiffusion}. This occurs whenever the function $W(k) \equiv i\omega(k)/k$ grows faster than linearly as $k \to 0$ along the real axis. In this final section, we investigate this possibility.

\subsection{Dispersion relation at real \texorpdfstring{$k$}{}}

Consider again the real-$k$ analysis of section \ref{longtermmemory}. Since the singularity of \eqref{psipsipsi} is integrable (recall equation \eqref{mildersingularity}), the velocity integral in \eqref{determinantequation} can be performed consistently, leading to the following implicit equation for $W(k)$:
\begin{equation}\label{roomM}
 \int_0^\infty d\varepsilon \,
 \dfrac{\varepsilon^2 e^{-\beta\varepsilon}}{\tau_\varepsilon}
 \left[
 1-
 \dfrac{1}{k\tau_\varepsilon}
 \arctan\!\left(
 \dfrac{k\tau_\varepsilon}{1-W k\tau_\varepsilon}
 \right)
 \right]
 =0 \, .
\end{equation}
We see that, although $\psi(\varepsilon,v^1)$ is singular when $1-Wk\tau_\varepsilon=0$, the full angle-integrated contribution (the term in the square bracket) remains finite, though discontinuous, since the arctangent approaches $\pm\pi/2$. This confirms that the singularity was indeed integrable. Hence, the eigenvalue condition remains well defined for real $k$, and we can safely use \eqref{roomM} to determine the dispersion relation of Fourier modes.

\subsection{Divergence of the diffusivity}

Equation \eqref{roomM} can be rearranged as an expression for $W$ as follows:
\begin{equation}\label{roomM2}
W= \dfrac{\displaystyle\int_0^\infty d(\beta\varepsilon) \, \dfrac{(\beta\varepsilon)^2 e^{-\beta\varepsilon}}{k\tau_\varepsilon} \left[1-\dfrac{1-Wk\tau_\varepsilon}{k\tau_\varepsilon} \arctan\!\left(\dfrac{k\tau_\varepsilon}{1-W k\tau_\varepsilon} \right)\right]}{\displaystyle\int_0^\infty d(\beta\varepsilon) \, \dfrac{(\beta\varepsilon)^2 e^{-\beta\varepsilon}}{k\tau_\varepsilon} \arctan\!\left(\dfrac{k\tau_\varepsilon}{1-W k\tau_\varepsilon} \right)} \, .
\end{equation}
For small $k$ and $W$, one may expand the integrand at fixed $\varepsilon$ using the expansion series, $\frac{1}{x}\text{arctan}(x)=\sum_{n=0}^{\infty}(-1)^n\frac{x^{2n}}{2n+1}$. Keeping terms up to the leading order, this yields the following expression for $W$,
\begin{equation}\label{roomM3}
W=\dfrac{k}{6} \int_0^\infty d(\beta\varepsilon) \, (\beta\varepsilon)^2 e^{-\beta\varepsilon} \tau_\varepsilon + \mathcal{O}(k^2) \, .
\end{equation}
Comparing with the diffusive form $W=\mathfrak{D}k$, one would identify\footnote{Note that, for power-law relaxation times $\tau_\varepsilon = \tau (\beta\varepsilon)^\gamma$, Eq.~\eqref{Diff} gives $\mathfrak{D} = \Gamma(3+\gamma)\tau/6$, in agreement with the hydrodynamic pole expansion obtained in momentum-dependent RTA correlation-function analyses~\cite{Brants:2024wrx}.}
\begin{equation}\label{Diff}
\mathfrak{D} = \dfrac{1}{6} \int_0^\infty d(\beta\varepsilon) \, (\beta\varepsilon)^2 e^{-\beta\varepsilon} \tau_\varepsilon \, .
\end{equation}
This coefficient diverges whenever $\tau_\varepsilon$ becomes sufficiently large in a region that is not suppressed strongly enough by the thermal weight. For instance, an infrared divergence arises if $\tau_\varepsilon \sim \varepsilon^{-3}$ as $\varepsilon \rightarrow 0$, while a sufficiently rapid growth such as $\tau_\varepsilon \sim  e^{\beta\varepsilon}$ would instead produce an ultraviolet divergence.

However, this divergence should be interpreted with care. The expansion leading to \eqref{roomM3} is performed at fixed $\varepsilon$, where it is always well-defined. Passing from the pointwise expansion to \eqref{roomM3} requires exchanging the small-$k$ limit with the $\varepsilon$-integration. This step is justified only if the $\mathcal{O}(k^2)$ remainder is uniformly integrable in $\varepsilon$ for sufficiently small $k$, which need not hold in gapless systems. When this condition fails, the integral of the subleading terms may be of the same order as the leading contribution, and the identification of $\mathfrak{D}$ as the coefficient of a linear expansion of $W(k)$ becomes invalid. In particular, the divergence of $\mathfrak{D}$ does not imply that $W(k)$ itself diverges, but rather that $W(k)$ grows faster than linearly in the limit $k \to 0$, meaning that the system fails to enter the ordinary diffusive universality class, and enters the superdiffusive universality class, which we study below.

\subsection{Superdiffusion}

Let us discuss the superdiffusive behavior in a concrete example. Consider
$\tau_\varepsilon=\tau/(\beta\varepsilon)^3$, for which \eqref{Diff} diverges. Defining $q=k\tau$ and changing integration variable to $\xi=\beta\varepsilon$, equation \eqref{roomM2} becomes,
\begin{equation}\label{roomM4}
W= \dfrac{\displaystyle\int_0^\infty d\xi \, \xi^5 e^{-\xi} \left[1-\dfrac{\xi^3-Wq}{q} \arctan\!\left(\dfrac{q}{\xi^3-W q} \right)\right]}{\displaystyle\int_0^\infty d\xi  \, \xi^5 e^{-\xi} \arctan\!\left(\dfrac{q}{\xi^3-W q} \right)} \, .
\end{equation}
To determine the leading behavior of $W(q)$ as $q\to 0$ (assuming $q>0$), note that the argument of the arctangent is small whenever $\xi^3 \gg q(1+W)$. We may therefore split the integrals at $\xi \sim [q(1+W)]^{1/3}$. 

In the region $\xi \lesssim [q(1+W)]^{1/3}$, the integration domain shrinks to zero as $q\to 0$, and the integrand remains bounded; hence this region gives a subleading contribution. In the complementary region $\xi \gtrsim [q(1+W)]^{1/3}$, we may expand the arctangent in powers of $q/\xi^3$, and neglect $Wq$ with respect to $\xi^3$, since $W\to 0$ as $q\to 0$. This yields
\begin{equation}\label{roomM5}
W\sim \dfrac{1}{3} \dfrac{\displaystyle\int_{q^{1/3}}^\infty d\xi \, \xi^5 e^{-\xi}  \left(\dfrac{q}{\xi^3-Wq} \right)^2}{\displaystyle\int_{q^{1/3}}^\infty d\xi  \, \xi^5 e^{-\xi}  \dfrac{q}{\xi^3-Wq}} \sim \dfrac{q}{3} \dfrac{\displaystyle\int_{q^{1/3}}^\infty d\xi \,  \dfrac{e^{-\xi}}{\xi}}{
\displaystyle\int_{q^{1/3}}^\infty d\xi  \, \xi^2 e^{-\xi}} \sim \dfrac{q}{18} \ln \left(\dfrac{1}{|q|} \right)\, .
\end{equation}
This confirms that the mode is superdiffusive, in the sense that $\omega(k)=-ikW(k)$ grows faster than $k^2$. Equivalently, the effective ratio $W(k)/k$ diverges logarithmically as $k\to 0$, even though $W(k)\to 0$.

This instance of superdiffusion is closely analogous to the well-known divergence of the diffusivity induced by hydrodynamic fluctuations in low spatial dimensions \cite{ErnstLongTimeTails1970,KovtunStickiness2011}. In that context, long-wavelength hydrodynamic modes have parametrically large lifetimes, and their nonlinear coupling leads to an infrared divergence of $\mathfrak{D}$ (notably in two dimensions). Physically, transport is dominated by fluctuations that decay so slowly that the system fails to coarse-grain to a local diffusive description.
In the present setting, the role of such modes is played by low-energy particles. For $\tau_\varepsilon=\tau/(\beta\varepsilon)^3$, the relaxation time diverges as $\varepsilon\to 0$, so that the system contains excitations with arbitrarily long equilibration timescales. As in fluctuating hydrodynamics, these slow degrees of freedom dominate the long-time dynamics and produce an infrared divergence of $\mathfrak{D}$. The analogy is structural: the divergence arises from an accumulation of modes with vanishing decay rate, with the infrared variable being momentum in the hydrodynamic case and energy in the present one. This signals the breakdown of diffusive scaling and the onset of superdiffusive behavior.

% \newpage
\section{Conclusions}

It was recently shown in \cite{GavassinoGapless:2024rck} that most realistic relativistic scattering cross sections lead to kinetic theories whose non-hydrodynamic sector is gapless. In such systems, there exist non-hydrodynamic excitations with arbitrarily long relaxation times, suggesting that microscopic memory may persist indefinitely. Nevertheless, one might still hope that an effectively hydrodynamic regime can be defined in practice, by restricting attention to sufficiently regular initial states. In QCD kinetic theory, for example, the long-lived modes are typically associated with high-energy particles that scatter inefficiently because the collision cross section decreases at large energies. One may therefore expect that hydrodynamics remains applicable whenever the initial distribution function does not strongly excite these sectors.

In this paper, we investigated obstructions to this program using a simple, analytically tractable kinetic model, namely an RTA theory with a momentum-dependent relaxation time. We found that, in gapless systems, any attempt to construct a purely hydrodynamic description inevitably encounters singular behavior precisely in the energy sectors where the long-lived non-hydrodynamic modes reside. In the RTA matching frame, this manifests itself in the divergence of the hydrodynamic gradient expansion on essentially all flows, including Fourier modes. At the spectral level, it appears in the fact that the hydrodynamic dispersion relation $\omega(k)$ is generically non-analytic at $k=0$, because the hydrodynamic excitation cannot be cleanly separated from the continuum of slow non-hydrodynamic modes. In sufficiently singular cases, even the leading transport coefficient (the diffusivity) may diverge, thereby making the effective long-time dynamics superdiffusive and intrinsically nonlocal. 

More generally, we found that any regularization of the singular hydrodynamic distribution function necessarily reintroduces non-hydrodynamic contributions that decay on the same timescale as the hydrodynamic mode itself. As a result, the system never fully forgets the non-hydrodynamic sector, making it impossible to sharply disentangle hydrodynamic and non-hydrodynamic dynamics through initialization alone.

We note that these effects are not necessarily dominant at finite times. Structurally, they resemble the long-time tails generated by hydrodynamic fluctuations \cite{KovtunStickiness2011}: their contribution may initially be parametrically small, yet it becomes increasingly important as faster modes decay away and only the longest-lived excitations survive. Eventually, the dynamics inevitably cross over toward dehydrodynamization~\cite{Kurkela:2017xis}.

\section*{Acknowledgements}

We thank Sayantani Bhattacharyya and Shuvayu Roy for fruitful discussions.
This work is supported by a MERAC Foundation prize grant,  an Isaac Newton Trust Grant, and funding from the Cambridge Centre for Theoretical Cosmology. R.S. is supported by a postdoctoral fellowship from the West University of Timisoara, Romania.

\bibliographystyle{utphys}
\bibliography{Biblio}
\end{document}